\documentclass{article}

\usepackage[final]{corl_2023} 
\usepackage{graphicx}
\usepackage{subcaption}
\usepackage{multirow}
\usepackage{wrapfig}
\usepackage{bm}

\usepackage{amsmath,amsfonts,bm}

\def\eqref#1{equation~\ref{#1}}

\def\1{\bm{1}}

\def\va{{\bm{a}}}

\DeclareMathAlphabet{\mathsfit}{\encodingdefault}{\sfdefault}{m}{sl}
\SetMathAlphabet{\mathsfit}{bold}{\encodingdefault}{\sfdefault}{bx}{n}

\def\gA{{\mathcal{A}}}

\def\gC{{\mathcal{C}}}

\def\gL{{\mathcal{L}}}

\def\gN{{\mathcal{N}}}

\def\gS{{\mathcal{S}}}
\def\gT{{\mathcal{T}}}

\title{Enhancing Multi-Agent Coordination through Common Operating Picture Integration}

\author{
  Peihong Yu \qquad Bhoram Lee \qquad Aswin Raghavan \qquad Supun Samarasekara\\
  SRI International\\
  \texttt{first.last@sri.com} \\
  \And
  Pratap Tokekar \\
  Unversity of Marlyand \\
  \texttt{tokekar@umd.edu} \\
  \And
  James Zachary Hare\\
  DEVCOM Army Research Laboratory\\
  \texttt{james.z.hare.civ@army.mil} \\
}

\begin{document}
\maketitle


\begin{abstract}
In multi-agent systems, agents possess only local observations of the environment. Communication between teammates becomes crucial for enhancing coordination. Past research has primarily focused on encoding local information into embedding messages which are unintelligible to humans. We find that using these messages in agent's policy learning leads to brittle policies when tested on out-of-distribution initial states.
We present an approach to multi-agent coordination, where each agent is equipped with the capability to integrate its (history of) observations, actions and messages received into a Common Operating Picture (COP) and disseminate the COP. This process takes into account the dynamic nature of the environment and the shared mission. 
We conducted experiments in the StarCraft2 environment to validate our approach. Our results demonstrate the efficacy of COP integration, and show that COP-based training leads to robust policies compared to state-of-the-art Multi-Agent Reinforcement Learning (MARL) methods when faced with out-of-distribution initial states. 
\end{abstract}



\section{Introduction}
We consider multi-agent tasks where agents are required to collaborate to accomplish a shared mission. Multi-Agent Reinforcement Learning (MARL) algorithms can learn decentralized policies, such that after training, each agent uses only its local observations. In such a setting, 
sharing information with each other can lead to better decision-making. In this paper, we present a technique that allows agents to \emph{learn} to communicate in a unified representation grounded in observable quantities. 


\begin{wrapfigure}{r}{0.45\textwidth}
\vspace{-10pt}
     \begin{center}
     \includegraphics[width=\linewidth]{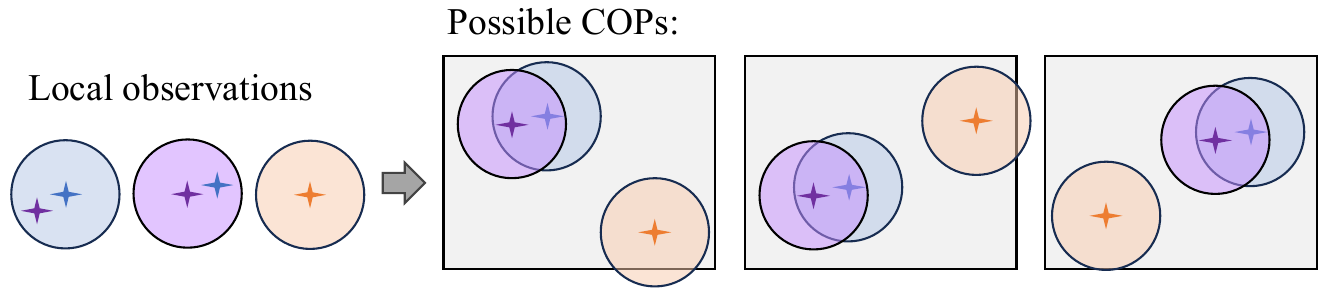}
     \end{center}
 \vspace{-5pt}
\caption{The purple and blue agents observe each other, the orange agent is beyond visual range. This leads to multiple potential COPs.} 
\label{fig:challenge-ex}
\end{wrapfigure}

Specifically, we equip each agent to integrate its (history of) own observations, actions and the communication messages into a  Common Operating Picture (COP) \cite{STEENTVEIT2021105381}. In this short paper, we consider the state of the underlying DecPOMDP as the COP. The COP contains the position and attributes of all the agents.
This COP is interpretable by human operators, making it valuable when considering human involvement in the decision-making process. We hypothesize that it improves the policy learning for each agent and essentially reduces the problem to single-agent RL. With better situational-awareness including non-local observations, we show that the COP leads to out-of-distribution generalization. 

There are several challenges in integrating observations and messages to form a COP in a decentralized fashion. 
    When each agent operates in an egocentric frame of reference, it is 
    difficult to integrate and align local observations (Figure \ref{fig:challenge-ex}).
    Agents are often constrained by a limited range of communication, introducing uncertainty into the COP-building process. This limitation can hinder the transmission of critical information.
    It is impractical or unsafe to communicate raw observations. 
    When agents communicate encoded COPs, the decoding process may introduce errors, potentially leading to hallucinations or incorrect interpretation of the communication. 

\begin{wrapfigure}{r}{0.4\textwidth}
\vspace{-10pt}
     \begin{center}
     \includegraphics[width=0.9\linewidth]{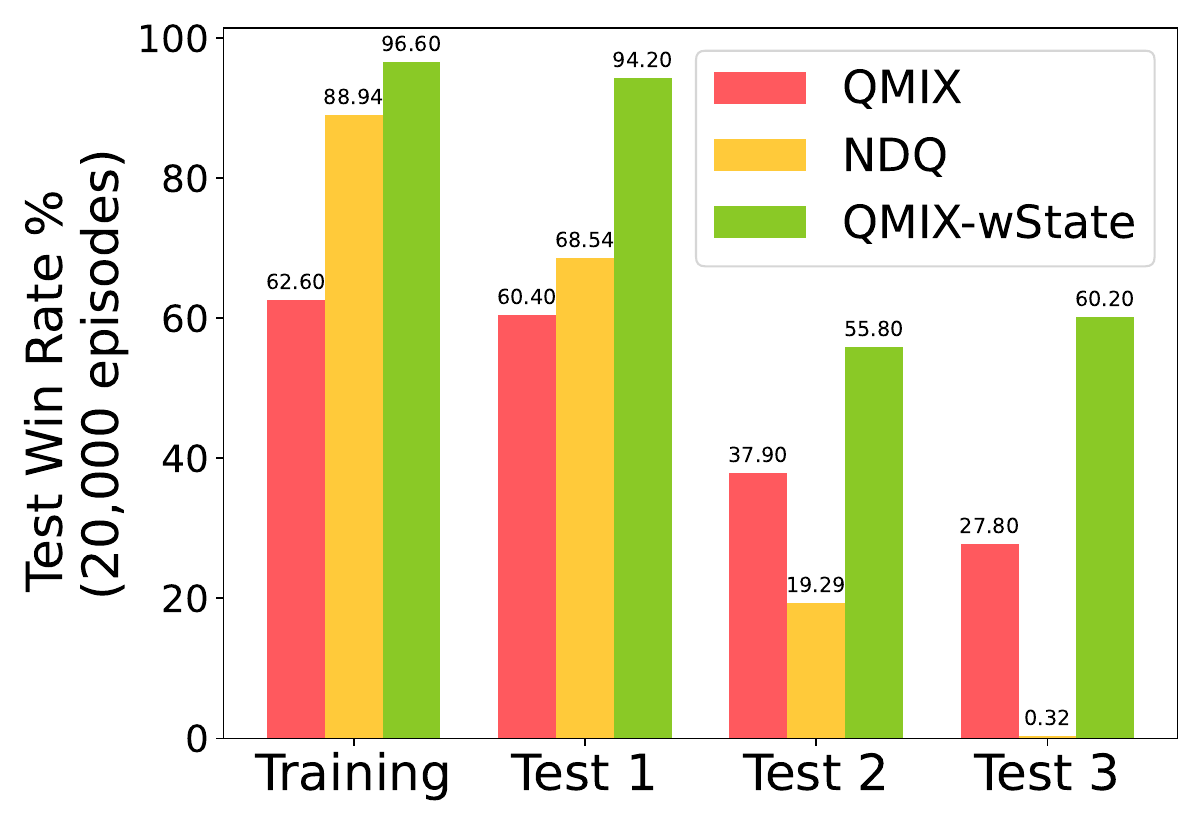}
     \end{center}
     \vspace{-5pt}
     \caption{Motivating COPs in MARL.} 
     \label{fig:motivation-exp}
\end{wrapfigure}

A quick experiment reveals that inferring such a COP will significantly benefit the QMIX \cite{rashid2018qmix} family of MARL algorithms. Within the Starcraft2 multi-agent challenge (SMAC) \cite{vinyals2017starcraft}, we tested QMIX on initial states (Test1-3) different from trained initial states. The experiment showed brittleness of  QMIX~\cite{rashid2018qmix} (without communication) and NDQ~\cite{wang2019learning} (with inter-agent communication). When we allow QMIX to access the state (from the simulator), the learned policy is less brittle as expected. This observation suggests that transmitting messages that help infer the global state can improve generalization in OOD scenarios. We present a MARL method that grounds the learned communication in the state.

\section{Method}

We consider the fully cooperative MARL problem with communication, which can be modeled as Decentralized Partially Observable Markov Decision Process (Dec-POMDP) \cite{oliehoek2016concise} and formulated as a tuple $\langle \gN, \gS, \gA, P, \Omega, O, R, \gamma, \gC \rangle$. The sets $\gN=\{1,...,n\}$ denotes the agents, $\gS$ are the states, $\gA$ are the actions, $\Omega$ are all observations, and $\gC$ are all possible communication messages. Each agent $i\in\gN$ acquires an observation $o_i \in \Omega, o_i=O(s, i), s\in\gS$. A joint action $\va = \langle a_1, ..., a_n \rangle$ leads to the next state $s' \sim P(s'|s, \va)$ and a shared global reward $r = R(s,\va)$.  

\begin{figure}[t]
    \begin{center}
    \includegraphics[width=0.9\linewidth]{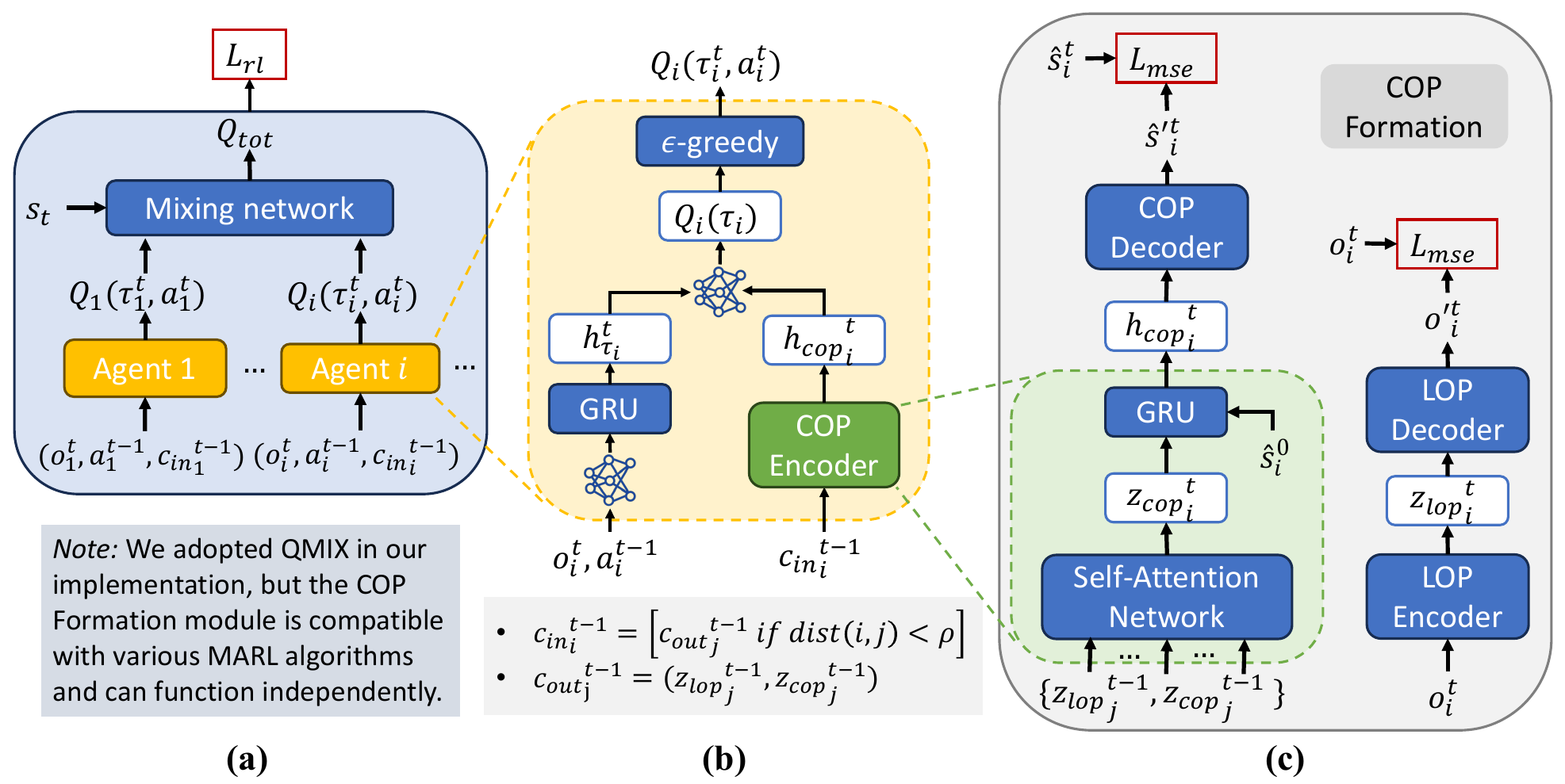}
    \end{center}
    \caption{The pipeline of our method. (a) The overall architecture. (b) Information integration and policy learning. (c) COP formation module.} 
    \label{fig:pipeline}
\end{figure}

Each agent selects action based on observation-action history $\tau_i \in \gT \equiv (\Omega \times \gA)^*$ using a policy $\pi(a_i|\tau_i, c^{in}_i)$. The policy is shared across agents during training.
Incoming communication for agent $i$, $c^{in}_i = [c^{out}_j \in \gC \text{ if } d(i, j)<\rho]$ is determined by the communication range  $\rho$ and distance between agents $d(i,j)$. The overall objective is to find a joint policy $\bm{\pi}(\bm{\tau}, \va)$ to maximize the global value function $Q_{\mathrm{tot}}^{\boldsymbol{\pi}}(\boldsymbol{\tau}, \boldsymbol{a})=\mathbb{E}_{s, \boldsymbol{a}}\left[\sum_{t=0}^{\infty} \gamma^t R(s, \boldsymbol{a}) \mid s_0=s, \boldsymbol{a}_{\mathbf{0}}=\boldsymbol{a}, \boldsymbol{\pi}\right]$, where $\bm{\tau}$ is the joint observation-action history and $\gamma \in [0,1)$ is the discount factor. We follow the \textit{Centralized Training and Decentralized Execution} (CTDE) paradigm and adopt QMIX's \cite{rashid2018qmix} architecture to build our algorithm, as illustrated in Figure \ref{fig:pipeline}.


\begin{wrapfigure}{r}{0.22\textwidth}
\vspace{-25pt}
    \begin{center}
    \includegraphics[width=0.95\linewidth]{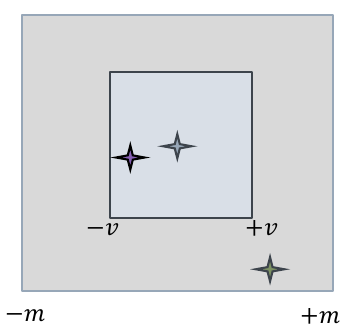}
    \end{center}
    \vspace{-5pt}
    \caption{Masking in the formation of egocentric COP. $v$ is the visual range of the agent and $m$ is the map size.} 
    \vspace{-10pt}
    \label{fig:ego-cop}
\end{wrapfigure}

We make a key assumption over prior work: agents have access to the initial state $s_0$. Our approach ``tracks'' the COP as a transformation of the initial state informed by observation-action history and incoming communication. We communicate two embeddings --- local observation-action embeddings and embeddings describing state evolution. The COP Formation module involves autoencoding these two key components:
(1) In the LOP (local operating picture) thread, we encode the local observation $o$ into an embedding denoted as $z_{lop}$ and then decode $o'$.
(2) In the COP thread, we encode all communicated messages using a Self-Attention network to deal with varying numbers of communication inputs, and obtain an embedding $z_{cop}$.

The state is tracked by GRU in hidden vector $h_{cop}$ using the initial state. Subsequently, $h_{cop}$ is decoded to produce the state $s'$. Note that $\hat{s}^0_i$, $\hat{s}^t_i$ are egocentric states centered on agent $i$. We use $\hat{s}^0_i$  to initialize the GRU. The final COP is produced by combining decoded LOP $o'$ and $s'$ as $\hat{s}'^t_i=\hat{s}'^t_i\times\mathbf{1}_{v_i}+o'^t_i\times(1-\mathbf{1}_{v_i})$, where $\mathbf{1}_{v_i}$ is a mask based on visual range of agent $i$. The masking is illustrated in Figure \ref{fig:ego-cop}. 

Our CTDE training objective comprises three parts $\gL = \gL_{RL} + \gL_{lop} + \gL_{cop}$, where (1) $\gL_{RL}$ is the Temporal Difference (TD)-based loss of QMIX, (2) $\gL_{lop} = L_{mse}(o', o)$ is the local observation reconstruction loss averaged over agents and time, and (3) $\gL_{cop}=L_{mse}(\hat{s}',\hat{s})$ is the egocentric state reconstruction loss averaged over agents and time. It is worth noting that COP formation can also be trained under a fixed policy. By producing an egocentric COP, the agent does not depend on knowledge of its own location in global frame. We produce an egocentric COP for each agent, and by comparing COPs across agents one could quantify uncertainty across agents. 


\section{Experiments}

\textbf{Methods}: We compare OOD generalization of our method against: (1) MASIA \cite{guan2022efficient} predicts the state from other agents' observations within a QMIX method, (2) NDQ \cite{wang2019learning} and TarMAC \cite{das2019tarmac} are prior work on MARL where the learned communication is not grounded on state prediction, (3) QMIX \cite{rashid2018qmix} and QMIX-based baselines QMIX\_wState and QMIX-Att where each agent has access to the state and other agents' observation, respectively. QMIX\_wState uses the state vector in addition to the agents' observation, a strong baseline for our method that assumes known initial state $s_0$. QMIX-Att uses an attention network to integrate others' observations, baseline for the self-attention architecture. Baseline architectures are shown in the appendix. 

\textbf{Scenarios}: We evaluate these methods on three maps from the Starcraft Multi-Agent Challenge \cite{vinyals2017starcraft} namely $1o\_10b\_vs\_1r$, $1o\_2r\_vs\_4r$, and $3s\_vs\_5z$. The first two maps involve an aerial agent to track and communicate the location of an enemy agent to other friendly agents. The third map is a micromanagement task. For each map, we created OOD initial states that are different than the training initial states. The OOD states are shown in the appendix. The methods are trained on the training map and evaluated in terms of win rate using episodes with the OOD initial states.

\begin{figure}[t]
\centering
\begin{subfigure}[c]{\columnwidth}
     \includegraphics[width=0.95\textwidth, keepaspectratio]{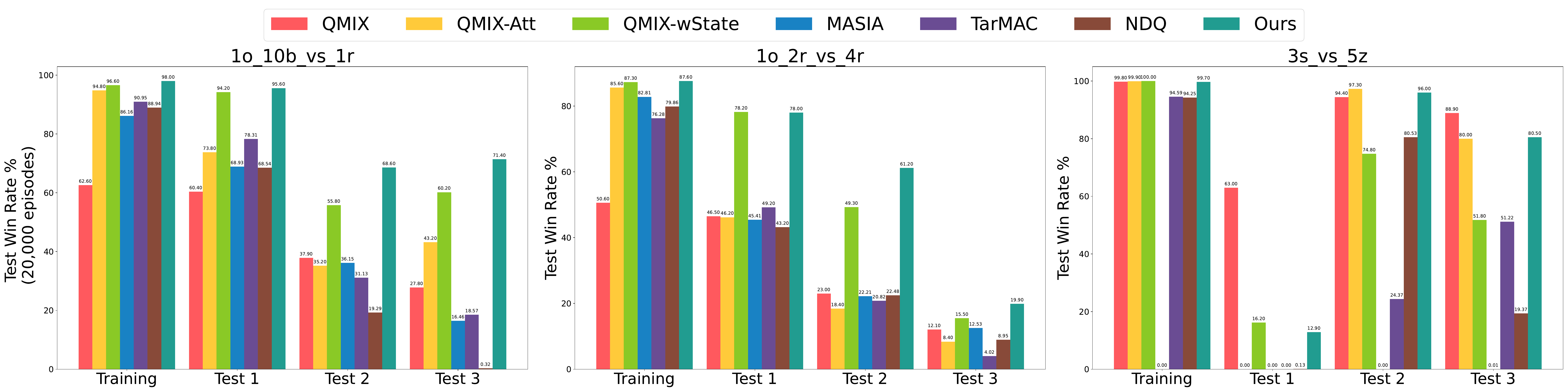}
\end{subfigure}
\caption{Comparison of methods on OOD initial states}
\label{fig:ood_generalization2}
\end{figure}

\textbf{Improved Generalization}: As shown in Figure \ref{fig:ood_generalization2}, our method achieves significantly higher win rates on all three maps  and in 8 out of 9 OOD states tested, compared to QMIX baselines and the prior MARL work. 
Compared to QMIX\_wState, the prior work MASIA, NDQ and TarMAC shows brittleness. Our method performs even better than QMIX\_wState, possibly due to training embeddings that  produce both winning policies and accurate state prediction. 

\begin{figure}[t]
\centering
\begin{subfigure}[c]{0.95\columnwidth}
     \includegraphics[width=\textwidth, keepaspectratio]{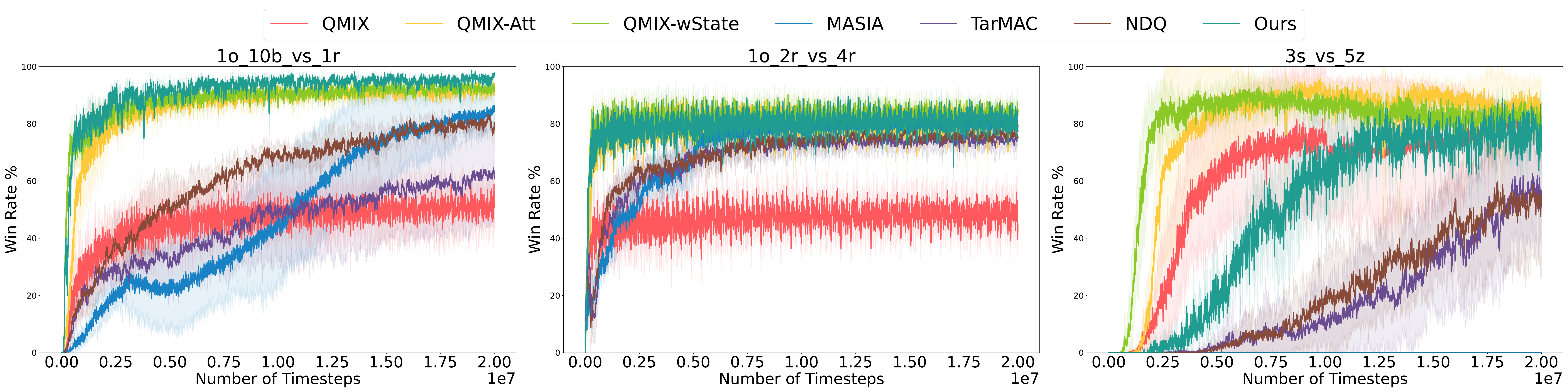}
\end{subfigure}
\caption{Training curves of various methods}
\label{fig:training_curves}
\end{figure}


\textbf{Sample Complexity}: We observe (Figure \ref{fig:training_curves}) that our method has significantly lower sample complexity and faster convergence on the maps that require communication beyond visual range ($1o\_10b\_vs\_1r$ and $1o\_2r\_vs\_4r$). Obviously, the performance of QMIX plateaus due to lack of communication, and MASIA and NDQ converge at a slower rate. These plots indicate that state prediction as in our COP-based training is an effective bias for quickly converging to robust policies. 

\textbf{Compression}: To understand the success of our method further, we reduce the embedding size from $32$ to $16, 8, $ then $4$ (Figure \ref{fig:emb}). At lower message length, the win rate on OOD maps is surprisingly only a bit lower. Even at message length of $4$, our method produces a significantly higher win rate than all the comparison methods. In the Appendix, we further ablate the impact of communication by dropping all messages in test episodes. 


\textbf{Evaluation of COPs}: Figure \ref{fig:sa} shows the average MSE of the produced COPs on training and OOD maps. We find that the COPs are highly accurate on training and the first two OOD maps (an MSE of $0.05$ is roughly only $6$ pixels in a field of view of $9$ pixels). We see clearly that the accuracy of the MSE (in our method) is directly correlated and predictive of win rate on OOD states. Beyond OOD generalization, our method produces human-interpretable COPs. Figure \ref{fig:ood-ex} shows an example COP where the location of the enemy is correctly predicted despite OOD initial state.



\begin{table} [ht]
\begin{minipage}{0.33\linewidth}
\centering
\includegraphics[width=\textwidth]{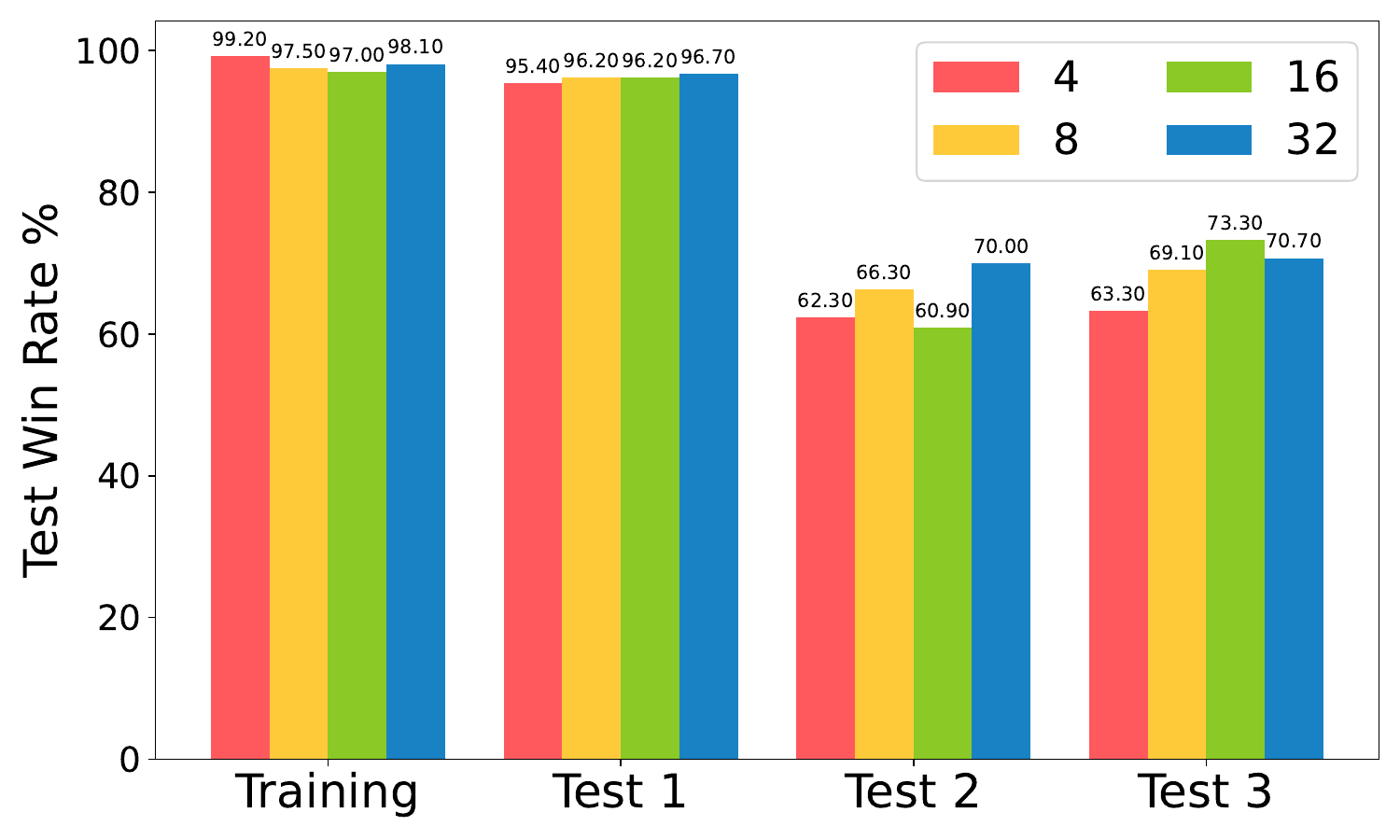}
\captionof{figure}{Varying message size ($1o\_10b\_vs\_1r$).}
\label{fig:emb}
\end{minipage}
\hfill
\begin{minipage}{0.3\linewidth}
\centering
\includegraphics[width=\textwidth]{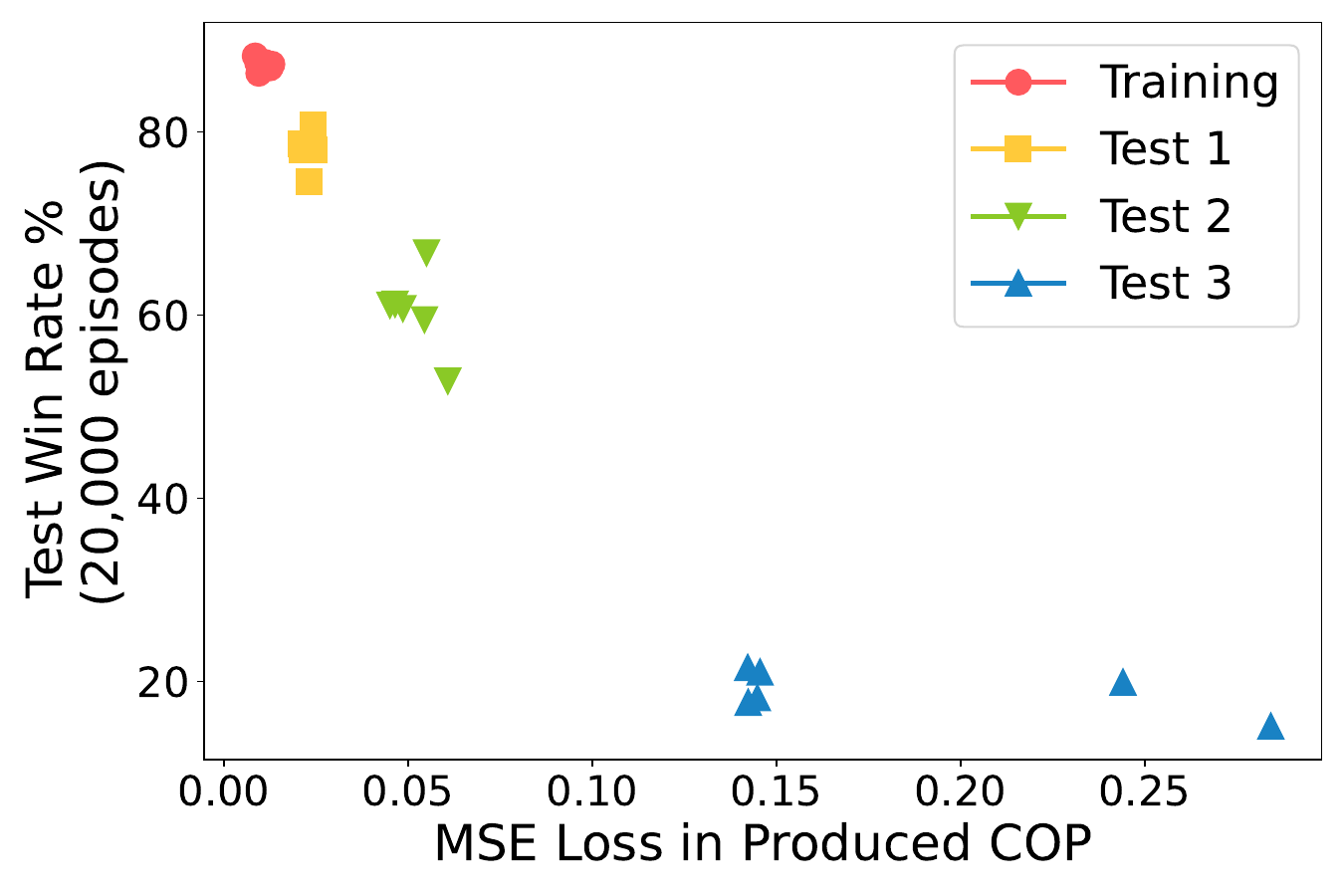}
\captionof{figure}{MSE vs win rate ($1o\_2r\_vs\_4r$).} 
\label{fig:sa}
\end{minipage}
\hfill
\begin{minipage}{0.33\linewidth}
\centering
\includegraphics[width=\textwidth]{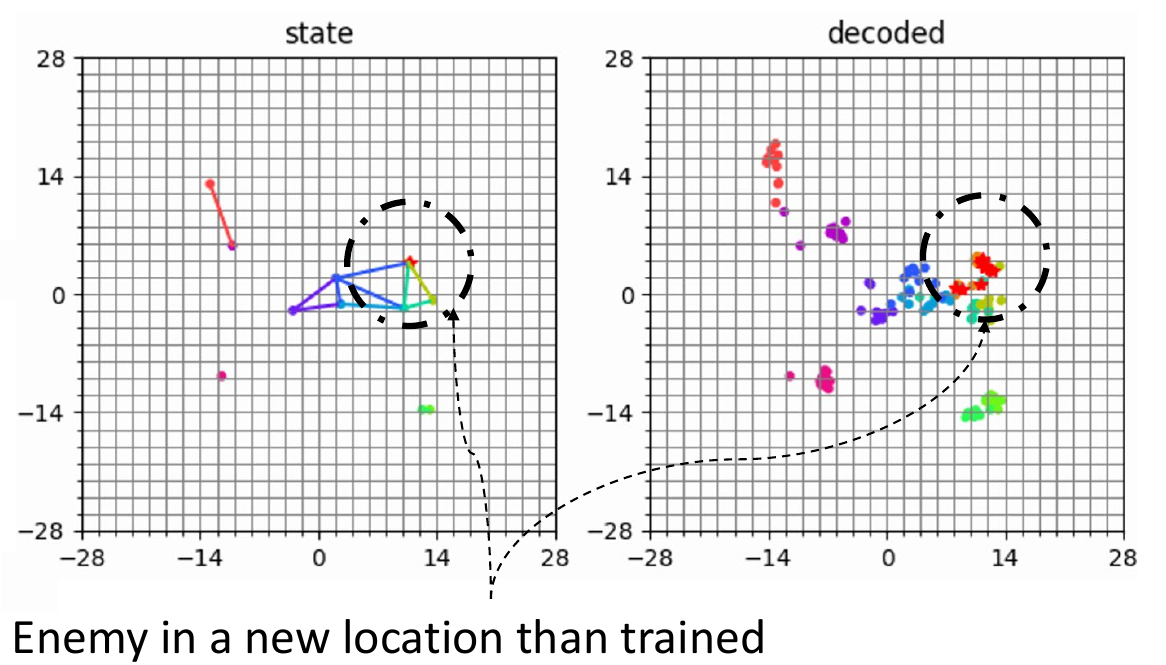}
\captionof{figure}{An example of reconstructed COP ($1o\_10b\_vs\_1r$).} 
\label{fig:ood-ex}
\end{minipage}
\end{table}

\acknowledgments 

This material is based upon work supported by the Army  Research Laboratory (ARL) under the Army AI Innovation Institute (A2I2) program Contract No. W911NF-22-2-0137. This work is supported in part by ONR under grant number N00014-18-1-2829.
\clearpage
\bibliography{example}  

\clearpage
\appendix
\section{Additional Experiments}

\subsection{Related Works}

Communication plays a vital role in multi-agent reinforcement Learning (MARL) systems \cite{zhu2022survey}, enabling agents to coordinate, share information, and collectively solve complex tasks. However, there remains an open question regarding what should be communicated and how to effectively utilize these communication messages. In various approaches, such as TarMAC \cite{das2019tarmac}, the sending agents generate both a signature and a message derived from their local observations. Meanwhile, the receiving agents employ attention mechanisms to combine and integrate these messages based on their relevance. Other methods, like NDQ \cite{wang2019learning}, introduce information-theoretic regularizers to minimize overall communication while maximizing the information conveyed by the messages for enhanced coordination. Some recent works explore more interpretable communications. For instance, \cite{lin2021learning} aims to establish a common understanding in communication symbols by autoencoding raw observations as the communication messages. Additionally, MASIA \cite{guan2022efficient} adopts an approach where raw observations are communicated and aggregated into a latent representation grounding the true state, aiming to reconstruct the global state from this embedding.

\subsection{Motivating Example in Details} Since the COP leads to better situational-awareness, we hypothesize that it also leads to improved out-of-distribution generalizations. We illustrate this with the following example. 
We use the $1o\_10b\_vs\_1r$ map from the Starcraft game~\cite{vinyals2017starcraft}. This map comprises a team consisting of 1 flying unit (Overseer) and 10 ground units (Banelings), with the collective objective of eliminating a single enemy unit (Roach). The flying and enemy units are initialized simultaneously at one of the four map corners. The ten ground units are distributed randomly across the map. 

\begin{figure}[ht]
\centering
\begin{subfigure}[c]{0.9\columnwidth}
     \includegraphics[width=\textwidth]{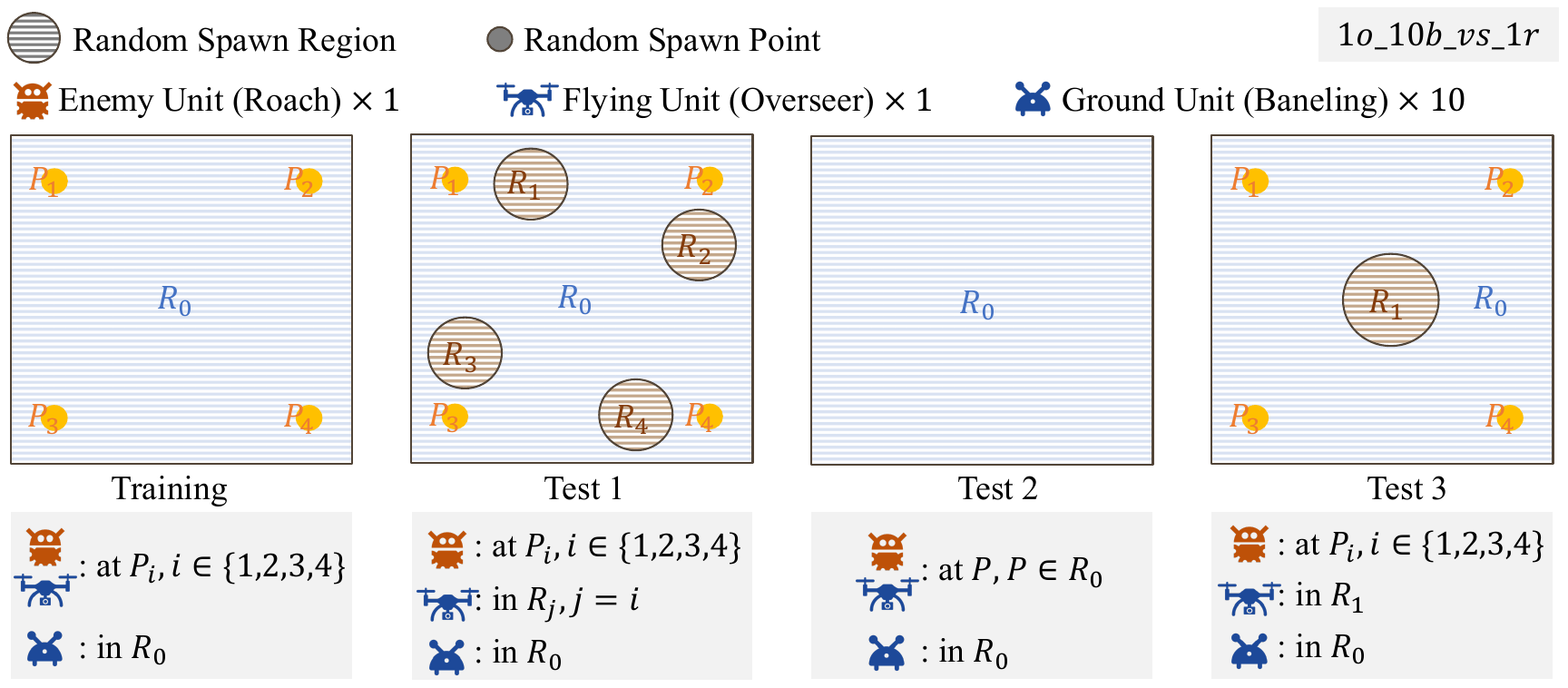}
     \caption{$1o\_10b\_vs\_1r$}
     \label{fig:ood-1}
\end{subfigure}
\begin{subfigure}[c]{0.9\columnwidth}
     \includegraphics[width=\textwidth]{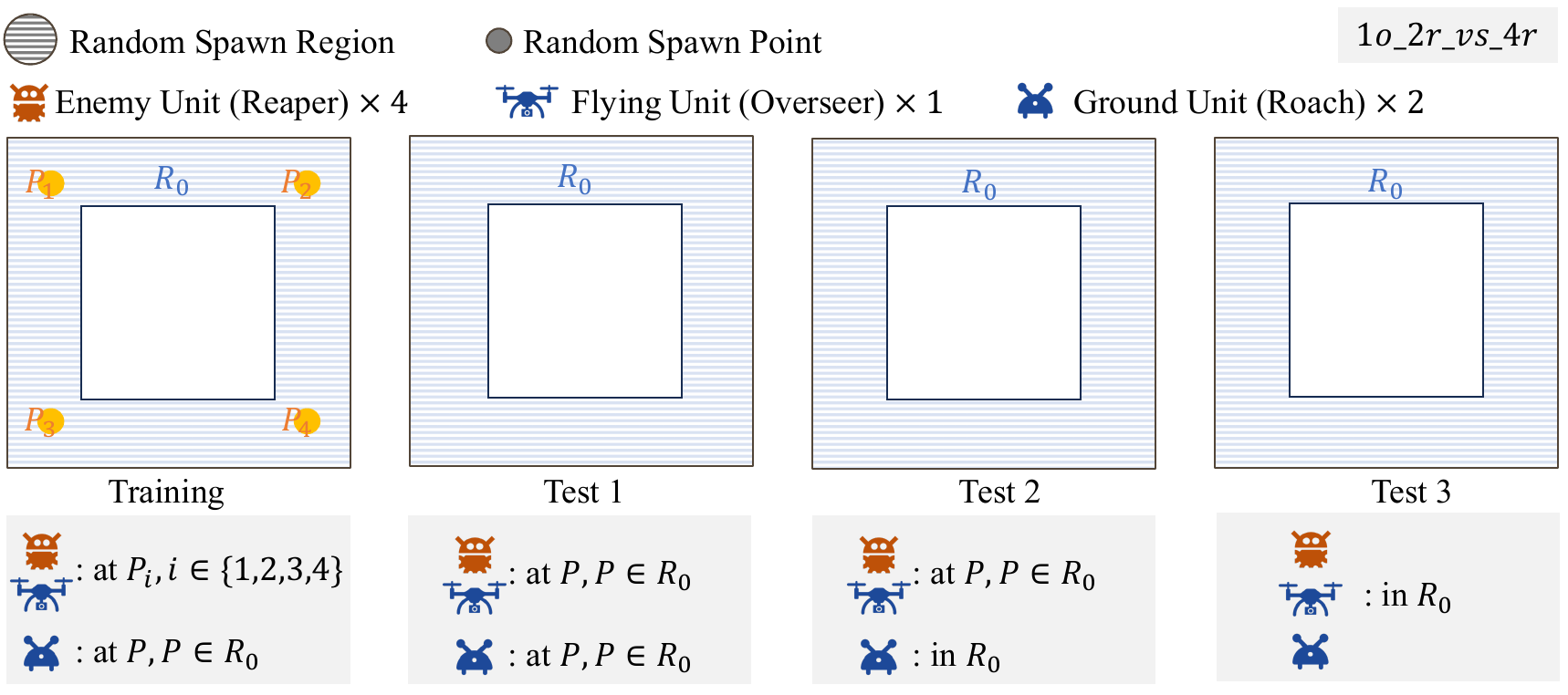}
     \caption{$1o\_2r\_vs\_4r$}
     \label{fig:ood-2}
\end{subfigure}
\begin{subfigure}[c]{0.9\columnwidth}
     \includegraphics[width=\textwidth]{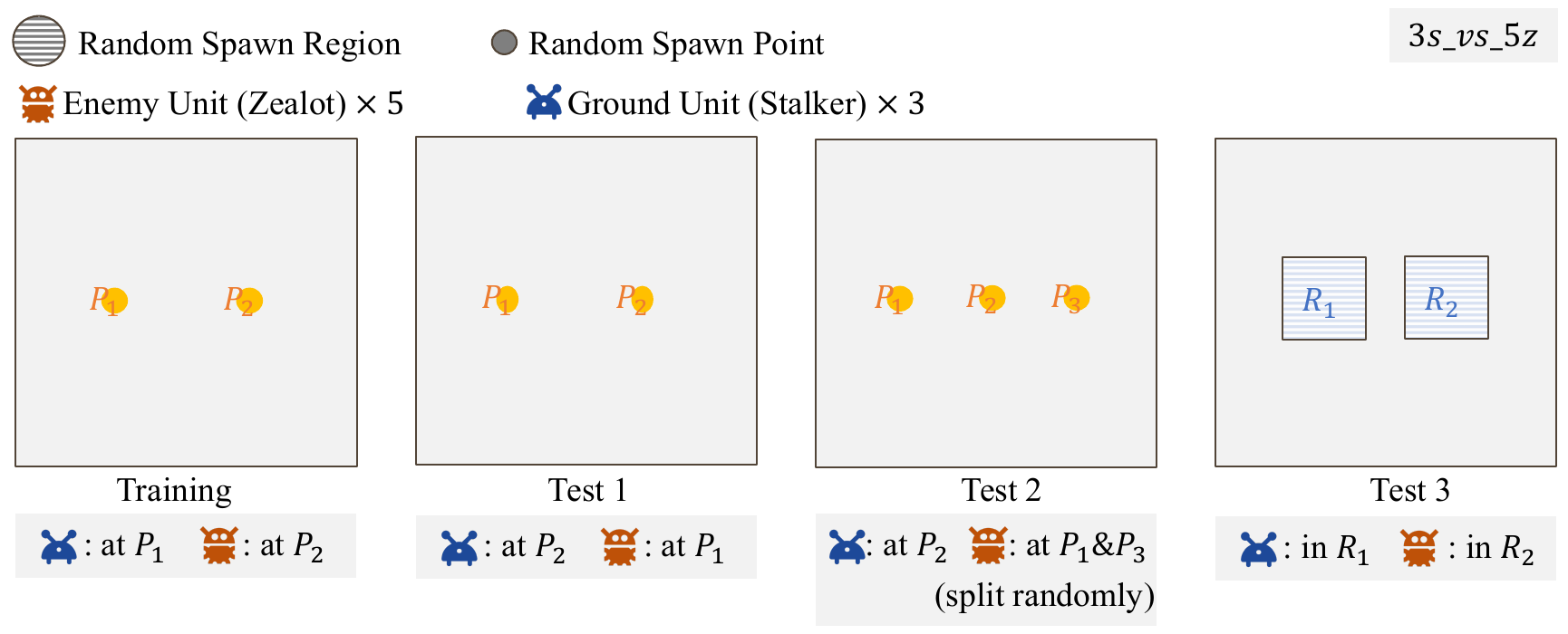}
     \caption{$3s\_vs\_5z$}
     \label{fig:ood-3}
\end{subfigure}
\caption{Illustration of modified initial distributions for $1o\_10b\_vs\_1r$, $1o\_2r\_vs\_4r$ and $3s\_vs\_5z$. In this figure, "at" means that all indicated agents are generated at the same point $P$, while "in" denotes that all indicated agents are randomly generated at distinct locations within the region $R$.}
\label{fig:ood}
\end{figure}

To introduce diversity and challenge, we have modified the initial distribution of the flying and enemy units and incorporated three out-of-distribution scenarios for this map, as shown in Figure \ref{fig:ood-1}. In Test 1, the setup closely resembles the training scenario. After selecting one corner to generate the enemy unit, we randomly position the flying unit within the nearby circle corresponding to that corner. Test 2 and Test 3 present greater challenges with notably different initial state distributions. In Test 2, flying and enemy units are generated together across the entire map, introducing a more complex spatial arrangement. In Test 3, the flying unit is generated inside the middle circle, considerably distant from the enemy unit, creating a scenario demanding different strategic considerations.

\begin{wrapfigure}{r}{0.5\textwidth}
    \begin{center}
    \includegraphics[width=0.8\linewidth]{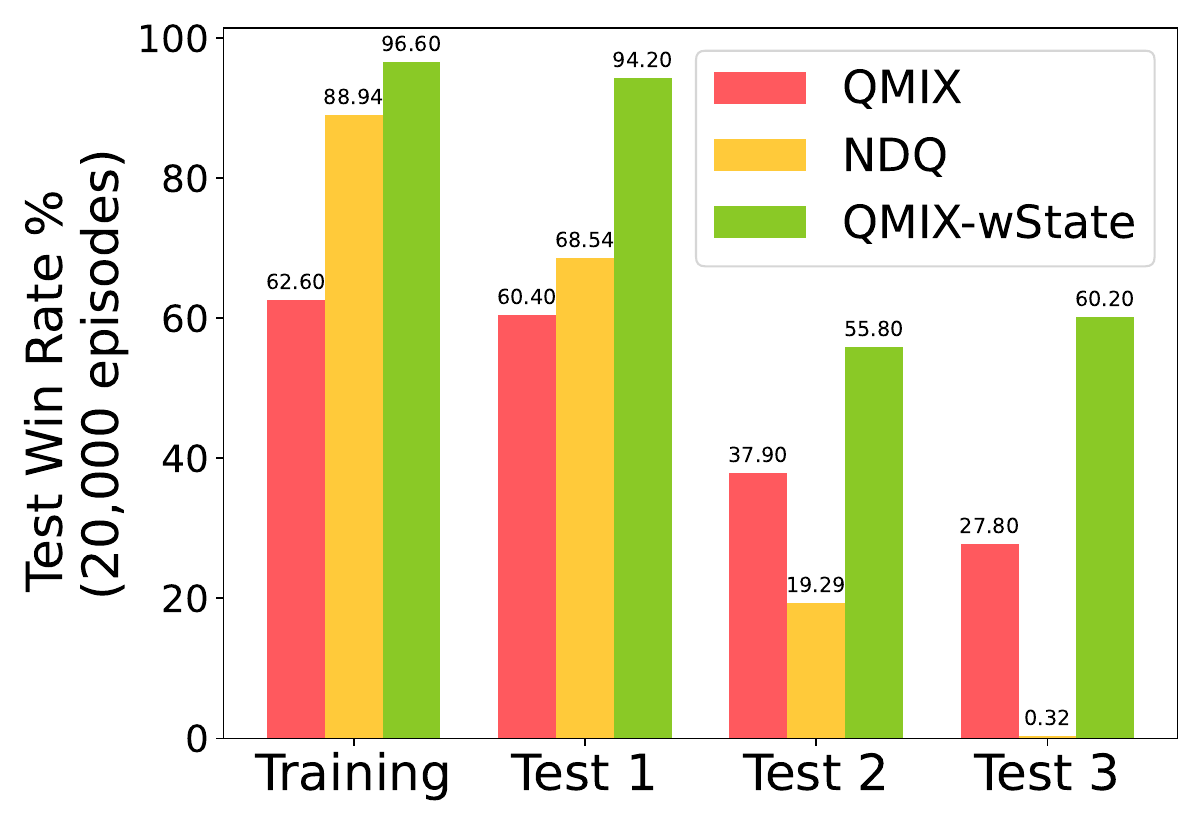}
    \end{center}
    \caption{Evaluation of MARL methods on alternate laydowns for $1o\_10b\_vs\_1r$.} 
    \label{fig:motivation-exp}
\end{wrapfigure}

We want to answer the following question: \textit{Will policies exhibit improved generalization on out-of-distribution (OOD) laydowns when agents possess a higher degree of situational awareness?} We employed two strong baseline algorithms to explore this question: QMIX~\cite{rashid2018qmix} (without communication) and NDQ~\cite{wang2019learning} (with inter-agent communication). Additionally, we implemented QMIX-wstate, which does not utilize communication but assumes that each agent has access to both its own partial observation and the global state. As shown in Figure \ref{fig:motivation-exp}, our findings indicate that QMIX-wstate outperforms the other algorithms across all OOD laydown scenarios. This observation strongly suggests that using the global state, i.e., COP,  is advantageous for enhancing generalization in OOD scenarios.

\newpage
\subsection{Visualization of OOD Laydowns}

\newpage
\subsection{Pipelines of QMIX-based Baselines}
\begin{figure}[ht]
\centering
\begin{subfigure}[c]{\columnwidth}
    \centering
     \includegraphics[width=0.6\textwidth]{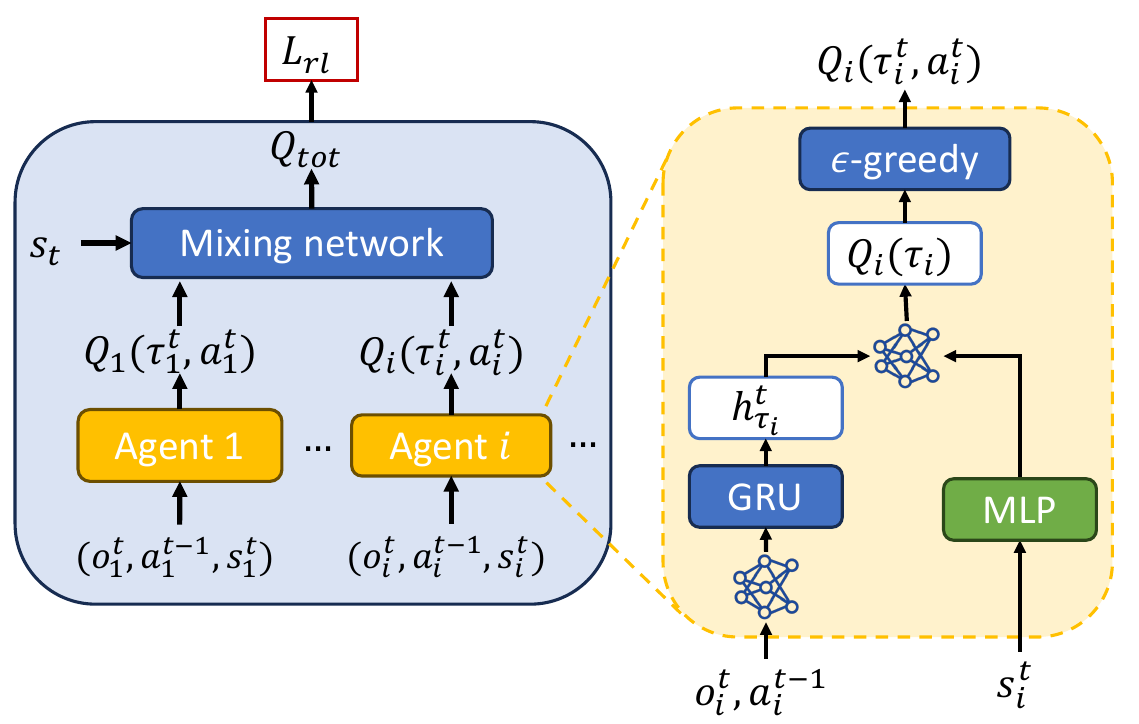}
     \caption{QMIX-wState: the agent takes $s_i^t$ as extra input at each time step.}
\end{subfigure}
\hspace{4mm}
\begin{subfigure}[c]{\columnwidth}
    \centering
     \includegraphics[width=0.65\textwidth]{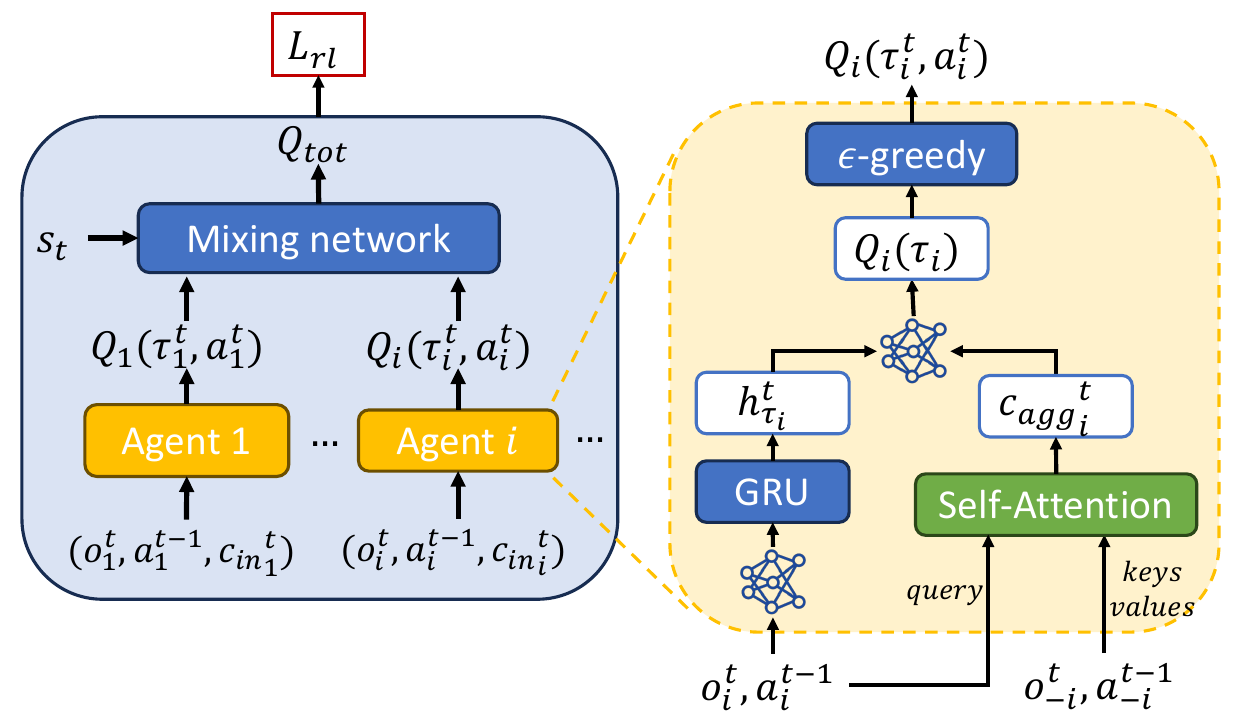}
     \caption{QMIX-Att: the agents receives all other agents' observation and action as communication messages.}
\end{subfigure}
\caption{Pipelines of QMIX-based baseline algorithms.}
\label{fig:qmix-pipeline}
\end{figure}

\subsection{Evaluate Message Dropping}


\begin{wrapfigure}{r}{0.45\textwidth}
\vspace{-20pt}
    \begin{center}
    \includegraphics[width=0.9\linewidth]{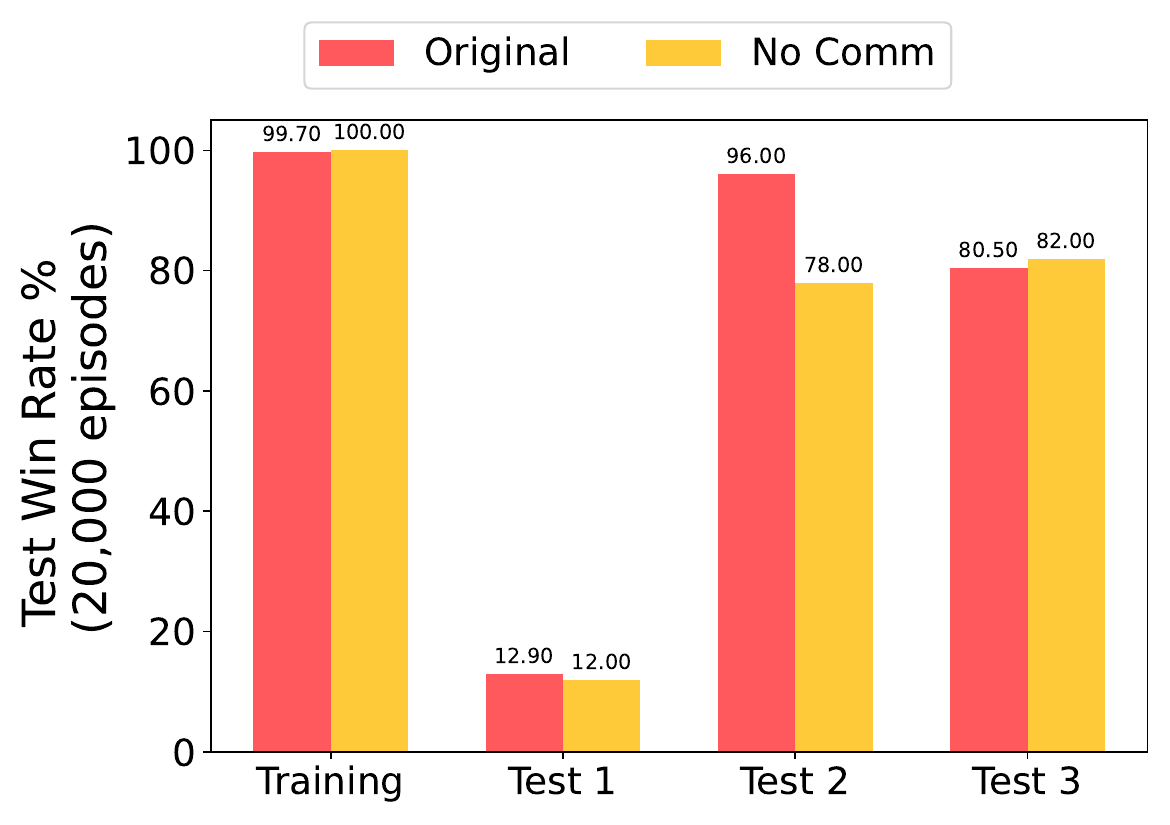}
    \end{center}
    \caption{Evaluation results of our algorithm with and without communication.}
    \label{fig:nocomm}
\end{wrapfigure}

To assess the significance of communication, we conducted experiments in which all messages were disabled during test episodes. We observe from $3s\_vs\_5z$ test 2 that the performance drops approximately $20\%$ in terms of win rate without communication. In other maps, the enemy is mostly stationary, so the information is primarily conveyed through the initial state and our approach effectively integrates this initial state information. However, even in such scenarios, we observed that the performance with communication was either equal to or superior to the performance without communication. This suggests that communication offers benefits beyond the initial state information, enhancing overall system performance.




\newpage
\subsection{QMIX with Varying Observation Sight Ranges}
Previous works \cite{guan2022efficient, shao2023complementary} discussed the dilemma in sight ranges: agents with small sight ranges can only observe limited information, making it challenging for them to engage in effective cooperation with their teammates. Conversely, agents with extensive sight ranges are more susceptible to distractions, degrading the cooperation quality. However, they only draw conclusions from the traffic junction and collaborative resource collection environments. 

We conduct similar experiments on the $1o\_10b\_vs\_1r$ map by running QMIX with varying observation sight ranges. Interestingly, we have observed that a larger sight range results in quicker convergence, an increased win rate, and improved generalization to out-of-distribution (OOD) scenarios. We posit that this phenomenon is intertwined with the characteristics of the SMAC environment. In this context, effective cooperation among allies is critical for locating and attacking enemies to secure victory, and having a broader perspective on the overall situation is essential for accomplishing these complex tasks.


\begin{figure}[ht]
\centering
\begin{subfigure}[c]{0.4\columnwidth}
     \includegraphics[width=\textwidth]{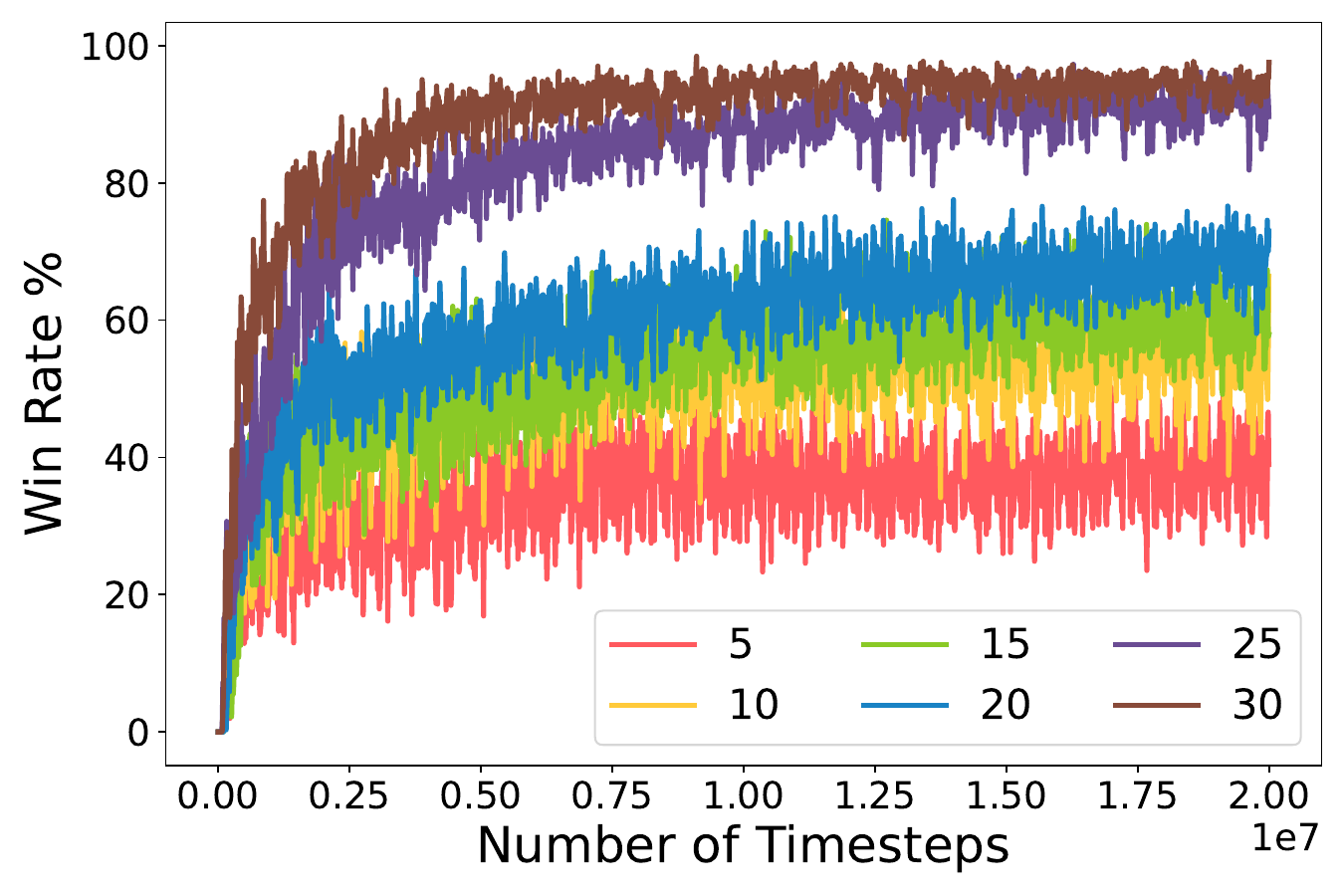}
     \caption{Training Curves}
\end{subfigure}
\hspace{0.025\columnwidth}
\begin{subfigure}[c]{0.53\columnwidth}
     \includegraphics[width=\textwidth]{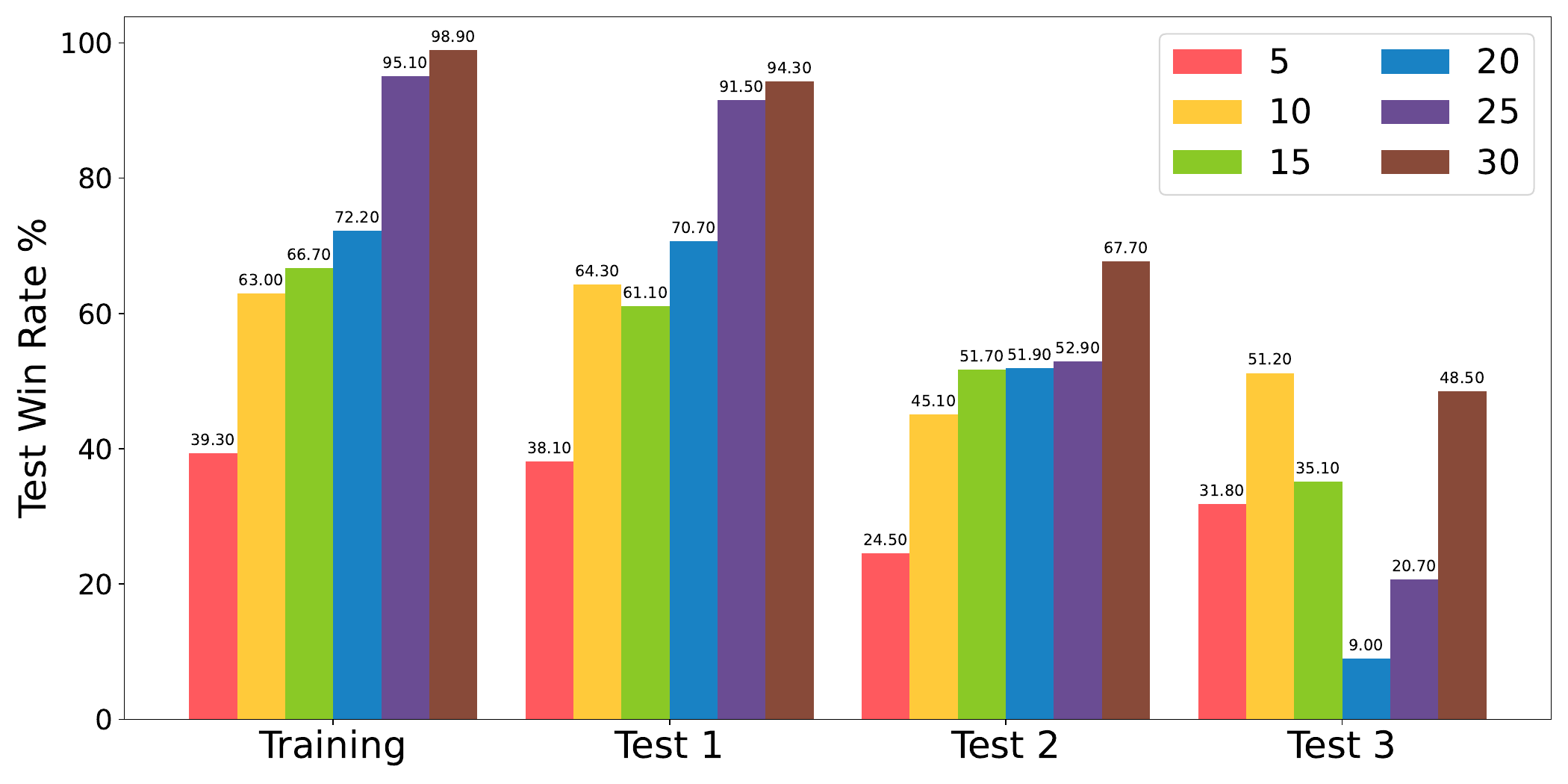}
     \caption{Evaluation on OOD Laydowns}
\end{subfigure}
\caption{Training and evaluation results of QMIX with different sight ranges on $1o\_10b\_vs\_1r$. A larger sight range gives us faster convergence speed and a higher win rate, which indicates that the agent can make wiser decisions if it is more aware of the global state under SMAC scenarios.}
\label{fig:sight_range}
\end{figure}

\begin{table}[htbp]
\centering
\begin{tabular}{|c|l|l|l|l|l|}
\hline
\textbf{Sight Range}         & \textbf{} & \textbf{Training} & \textbf{Test 1} & \textbf{Test 2} & \textbf{Test 3} \\ \hline
\multirow{2}{*}{\textbf{5}}  & Win Rate  & 39.30\%           & 38.10\%         & 24.50\%         & 31.80\%         \\ \cline{2-6} 
                             & Return    & $9.74\pm8.77$     & $9.52\pm8.73$   & $7.45\pm7.61$   & $8.65\pm8.30$   \\ \hline
\multirow{2}{*}{\textbf{10}} & Win Rate  & 63.00\%           & 64.30\%         & 45.10\%         & 51.20\%         \\ \cline{2-6} 
                             & Return    & $14.12\pm8.42$    & $14.40\pm8.30$  & $11.13\pm8.62$  & $12.34\pm8.51$  \\ \hline
\multirow{2}{*}{\textbf{15}} & Win Rate  & 66.70\%           & 61.10\%         & 51.70\%         & 35.10\%         \\ \cline{2-6} 
                             & Return    & $15.50\pm7.17$    & $14.73\pm7.30$  & $13.38\pm7.46$  & $9.86\pm8.02$   \\ \hline
\multirow{2}{*}{\textbf{20}} & Win Rate  & 72.20\%           & 70.70\%         & 51.90\%         & 9.00\%          \\ \cline{2-6} 
                             & Return    & $16.59\pm6.28$    & $16.37\pm6.40$  & $13.78\pm7.00$  & $6.53\pm4.77$   \\ \hline
\multirow{2}{*}{\textbf{25}} & Win Rate  & 95.10\%           & 91.50\%         & 52.90\%         & 20.70\%         \\ \cline{2-6} 
                             & Return    & $19.74\pm2.84$    & $19.26\pm3.66$  & $13.86\pm6.95$  & $8.55\pm6.32$   \\ \hline
\multirow{2}{*}{\textbf{30}} & Win Rate  & 98.90\%           & 94.30\%         & 67.70\%         & 48.50\%         \\ \cline{2-6} 
                             & Return    & $20.21\pm1.32$    & $19.62\pm3.01$  & $16.04\pm6.25$  & $13.28\pm6.99$  \\ \hline
\end{tabular}
\caption{Evaluations results on $1o\_10b\_vs\_1r$ over 20,000 episodes.}
\end{table}

\newpage
\subsection{Addtional Evaluation Details}

\begin{table}[ht]
\centering
\begin{tabular}{|c|l|l|l|l|l|}
\hline
\textbf{\begin{tabular}[c]{@{}c@{}}Embedding\\ Size\end{tabular}} &          & \textbf{Training} & \textbf{Test 1} & \textbf{Test 2} & \textbf{Test 3} \\ \hline
\multirow{2}{*}{\textbf{4}}                                       & Win Rate & 99.20\%           & 95.40\%         & 62.30\%         & 63.30\%         \\ \cline{2-6} 
                                                                  & Return   & $20.26\pm1.14$    & $19.77\pm2.73$  & $14.85\pm7.22$  & $15.21\pm6.92$  \\ \hline
\multirow{2}{*}{\textbf{8}}                                       & Win Rate & 97.50\%           & 96.20\%         & 66.30\%         & 69.10\%         \\ \cline{2-6} 
                                                                  & Return   & $20.04\pm1.98$    & $19.87\pm2.43$  & $15.48\pm6.93$  & $16.25\pm6.23$  \\ \hline
\multirow{2}{*}{\textbf{16}}                                      & Win Rate & 97.00\%           & 96.20\%         & 60.90\%         & 73.30\%         \\ \cline{2-6} 
                                                                  & Return   & $19.98\pm2.19$    & $19.89\pm2.44$  & $14.54\pm7.40$  & $16.74\pm6.12$  \\ \hline
\multirow{2}{*}{\textbf{32}}                                      & Win Rate & 98.10\%           & 96.70\%         & 70.00\%         & 70.70\%         \\ \cline{2-6} 
                                                                  & Return   & $20.10\pm1.75$    & $19.93\pm2.27$  & $16.11\pm6.56$  & $16.47\pm6.14$  \\ \hline
\end{tabular}
\caption{Comparison of our algorithm with different embedding sizes on $1o\_10b\_vs\_1r$ over 20,000 episodes.}
\end{table}

\begin{table}[ht]
\begin{tabular}{|c|l|l|l|l|l|}
\hline
\textbf{}                                         &                & \textbf{Training}       & \textbf{Test 1}         & \textbf{Test 2}         & \textbf{Test 3}         \\ \hline
\multirow{2}{*}{\textbf{QMIX}}                    & Win Rate   & 62.60\%                 & 60.40\%                 & 37.90\%                 & 27.80\%                 \\ \cline{2-6} 
                                                  & Return & $13.88\pm8.64$          & $13.59\pm8.63$          & $9.78\pm8.52$           & $8.73\pm7.62$           \\ \hline
\multirow{2}{*}{\textbf{QMIX-Attn}}               & Win Rate   & 94.80\%                 & 73.80\%                 & 35.20\%                 & 43.20\%                 \\ \cline{2-6} 
                                                  & Return & $19.69\pm2.93$          & $16.23\pm7.10$          & $9.63\pm8.15$           & $11.19\pm8.28$          \\ \hline
\multirow{2}{*}{\textbf{QMIX-wState}}             & Win Rate   & 96.60\%                 & 94.20\%                 & 55.80\%                 & 60.20\%                 \\ \cline{2-6} 
                                                  & Return & $19.92\pm2.39$          & $19.60\pm3.12$          & $13.31\pm8.11$          & $14.61\pm7.27$          \\ \hline
\multirow{2}{*}{\textbf{MASIA}}                   & Win Rate   & 86.16\%                 & 68.93\%                 & 36.15\%                 & 16.46\%                 \\ \cline{2-6} 
                                                  & Return & $18.56\pm4.56$          & $15.59\pm7.34$          & $9.86\pm8.16$           & $6.7798\pm6.50$         \\ \hline
\multirow{2}{*}{\textbf{TarMAC}}                  & Win Rate   & 90.95\%                 & 78.31\%                 & 31.13\%                 & 18.57\%                 \\ \cline{2-6} 
                                                  & Return & $19.25\pm3.81$          & $17.14\pm6.42$          & $8.93\pm7.97$           & $6.63\pm7.00$           \\ \hline
\multirow{2}{*}{\textbf{NDQ}}                     & Win Rate   & 88.94\%                 & 68.54\%                 & 19.29\%                 & 0.32\%                  \\ \cline{2-6} 
                                                  & Return & $18.93\pm4.18$          & $15.74\pm7.01$          & $6.96\pm6.99$           & $1.68\pm2.22$           \\ \hline
\multirow{2}{*}{\textbf{Our Method}} & Win Rate   & \textbf{98.00\%}        & \textbf{95.60\%}        & \textbf{68.60\%}        & \textbf{71.40\%}        \\ \cline{2-6} 
                                                  & Return & \bm{$20.11\pm1.77$} & \bm{$19.81\pm2.63$} & \bm{$15.79\pm6.90$} & \bm{$16.42\pm6.37$} \\ \hline
\end{tabular}
\caption{Evaluation results of various algorithms on $1o\_10b\_vs\_1r$ over 20,000 episodes.}
\end{table}

\begin{table}[ht]
\centering
\begin{tabular}{|c|l|l|l|l|l|}
\hline
\textbf{}                             &          & \textbf{Training}       & \textbf{Test 1}         & \textbf{Test 2}         & \textbf{Test 3}        \\ \hline
\multirow{2}{*}{\textbf{QMIX}}        & Win Rate & 50.60\%                 & 46.50\%                 & 23.00\%                 & 12.10\%                \\ \cline{2-6} 
                                      & Return   & $10.75\pm9.58$          & $9.86\pm9.65$           & $7.74\pm7.46$           & $7.80\pm5.15$          \\ \hline
\multirow{2}{*}{\textbf{QMIX-Attn}}   & Win Rate & 85.60\%                 & 46.20\%                 & 18.40\%                 & 8.40\%                 \\ \cline{2-6} 
                                      & Return   & $18.27\pm4.28$          & $9.81\pm9.64$           & $7.09\pm6.91$           & $7.10\pm4.68$          \\ \hline
\multirow{2}{*}{\textbf{QMIX-wState}} & Win Rate & 87.30\%                 & \textbf{78.20\%}        & 49.30\%                 & 15.50\%                \\ \cline{2-6} 
                                      & Return   & $18.51\pm3.96$          & \bm{$16.82\pm6.28$} & $13.18\pm7.01$          & $8.78\pm5.30$          \\ \hline
\multirow{2}{*}{\textbf{MASIA}}       & Win Rate & 82.81\%                & 45.41\%                  & 22.21\%             & 12.53\%               \\ \cline{2-6} 
                                      & Return   & $17.93\pm4.64$         & $9.51\pm9.73$            & $7.19\pm7.51$       & $7.39\pm5.43$         \\ \hline
\multirow{2}{*}{\textbf{TarMAC}}      & Win Rate & 76.28\%                 & 49.20\%                 & 20.82\%                 & 4.02\%                 \\ \cline{2-6} 
                                      & Return   & $16.60\pm6.34$          & $10.56\pm9.53$          & $7.11\pm7.29$           & $5.79\pm3.96$          \\ \hline
\multirow{2}{*}{\textbf{NDQ}}         & Win Rate & 79.86\%                 & 43.20\%                 & 22.48\%                 & 8.95\%                 \\ \cline{2-6} 
                                      & Return   & $17.56\pm4.97$          & $9.28\pm9.57$           & $7.36\pm7.46$           & $6.72\pm4.95$          \\ \hline
\multirow{2}{*}{\textbf{Our Method}}  & Win Rate & \textbf{87.60\%}        & 78.00\%                 & \textbf{61.20\%}        & \textbf{19.90\%}       \\ \cline{2-6} 
                                      & Return   & \bm{$18.61\pm3.75$} & $16.71\pm6.50$          & \bm{$14.92\pm6.70$} & \bm{$9.61\pm5.54$} \\ \hline
\end{tabular}
\caption{Evaluation results of various algorithms on $1o\_2r\_vs\_4r$ over 20,000 episodes.}
\end{table}

\begin{table}[ht]
\centering
\begin{tabular}{|c|l|l|l|l|l|}
\hline
\textbf{}                             &          & \textbf{Training}       & \textbf{Test 1}                & \textbf{Test 2}                & \textbf{Test 3}                \\ \hline
\multirow{2}{*}{\textbf{QMIX}}        & Win Rate & 99.80\%                 & \textbf{63.00\%}               & 94.40\%                        & \textbf{88.90\%}               \\ \cline{2-6} 
                                      & Return   & $21.17\pm0.62$          & \bm{$20.03\pm4.05$}        & $21.66\pm1.54$                 & \bm{$20.86\pm2.50$}        \\ \hline
\multirow{2}{*}{\textbf{QMIX-Attn}}   & Win Rate & 99.90\%                 & 0.00\%                         & \textbf{97.30\%}               & 80.00\%                        \\ \cline{2-6} 
                                      & Return   & $22.62\pm0.44$          & $0.01\pm0.23$                  & \bm{$22.43\pm1.30$}        & $20.94\pm3.67$                 \\ \hline
\multirow{2}{*}{\textbf{QMIX-wState}} & Win Rate & \textbf{100.00\%}       & 16.20\%                        & 74.80\%                        & 51.80\%                        \\ \cline{2-6} 
                                      & Return   & \bm{$25.26\pm1.13$} & $18.09\pm7.63$                 & $24.89\pm3.57$                 & $17.39\pm9.64$                 \\ \hline
\multirow{2}{*}{\textbf{MASIA}}       & Win Rate & 0.00\%                  & 0.00\%                         & 0.00\%                         & 0.01\%                         \\ \cline{2-6} 
                                      & Return   & $9.81\pm0.39$           & $5.91\pm2.08$                  & $5.28\pm1.74$                  & $7.73\pm1.41$                 \\ \hline
\multirow{2}{*}{\textbf{TarMAC}}      & Win Rate & 94.59\%                 & 0.00\%                         & 24.37\%                        & 51.22\%                        \\ \cline{2-6} 
                                      & Return   & $21.67\pm1.29$          & $0.00\pm0.11$                  & $18.84\pm4.14$                 & $19.30\pm4.60$                 \\ \hline
\multirow{2}{*}{\textbf{NDQ}}         & Win Rate & 94.25\%                 & 0.13\%                         & 80.53\%                        & 19.37\%                        \\ \cline{2-6} 
                                      & Return   & $22.08\pm1.50$          & $6.10\pm2.59$                  & $21.72\pm2.89$                 & $14.99\pm5.23$                 \\ \hline
\multirow{2}{*}{\textbf{Our Method}}  & Win Rate & 99.70\%                 & 12.90\%                        & 96.00\%                        &   80.50\%                 \\ \cline{2-6} 
                                      & Return   & $20.95\pm0.60$          & $12.37\pm7.28$                 & $21.22\pm1.35$                 & $20.52\pm3.26$                \\ \hline
\end{tabular}
\caption{Evaluation results of various algorithms on $3s\_vs\_5z$ over 20,000 episodes.}
\end{table}

\newpage
\subsection{Hallucination}
We also investigate the occurrence of false positives (hallucinations) at different time steps in the "$1o\_10b\_vs\_1r$" scenario. By employing the disentangle loss to enforce local consistency, we have observed a noteworthy outcome: no hallucinations transpire within the agents' field of view. The instances of hallucinations that do occur outside the field of view can be primarily attributed to agents that are no longer active or are considered "dead." These hallucinations tend to persist for a few time steps before fading away.

\begin{figure}[ht]
\centering
\begin{subfigure}[c]{0.24\columnwidth}
     \includegraphics[width=\textwidth]{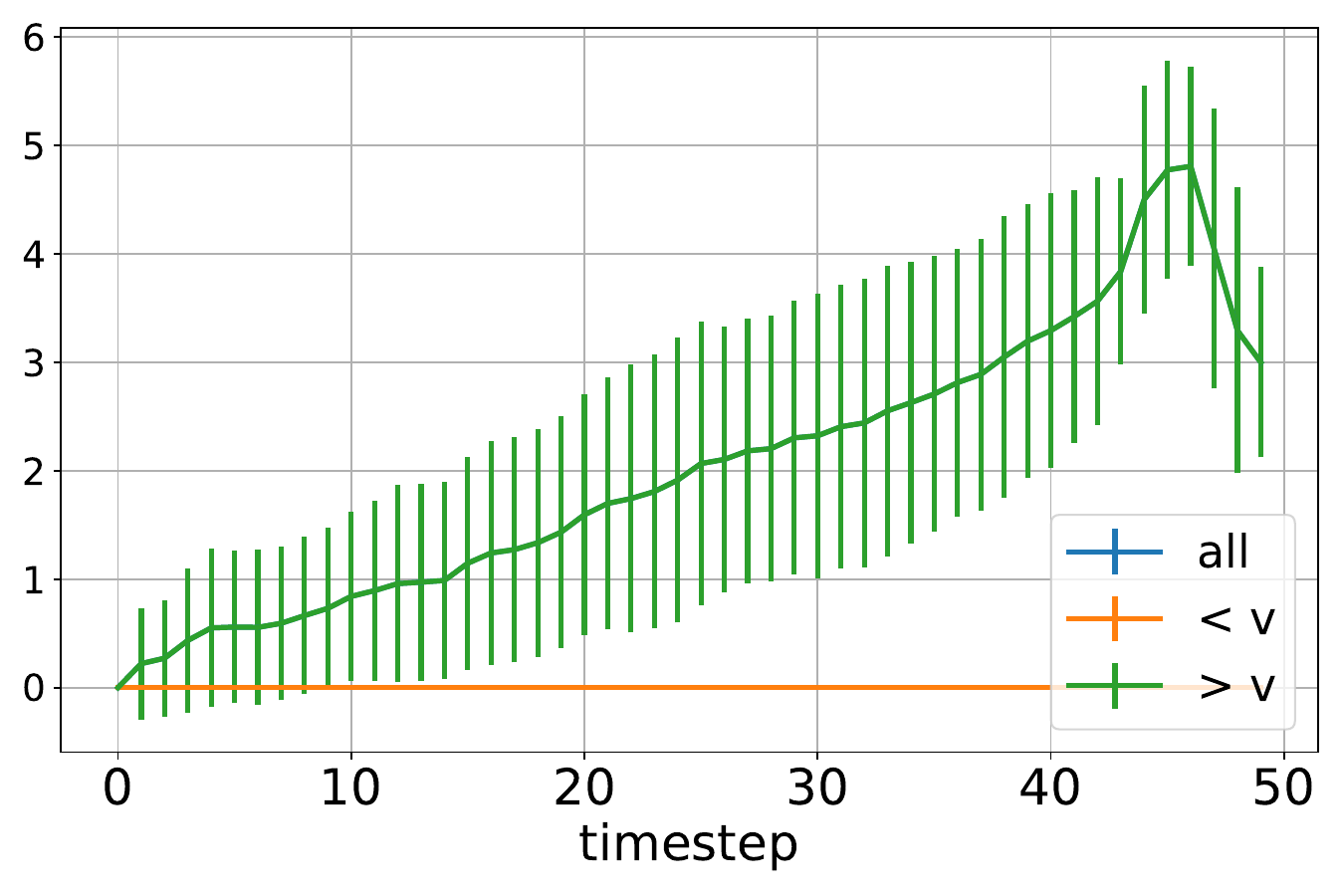}
     \caption{Training}
\end{subfigure}
\begin{subfigure}[c]{0.24\columnwidth}
     \includegraphics[width=\textwidth]{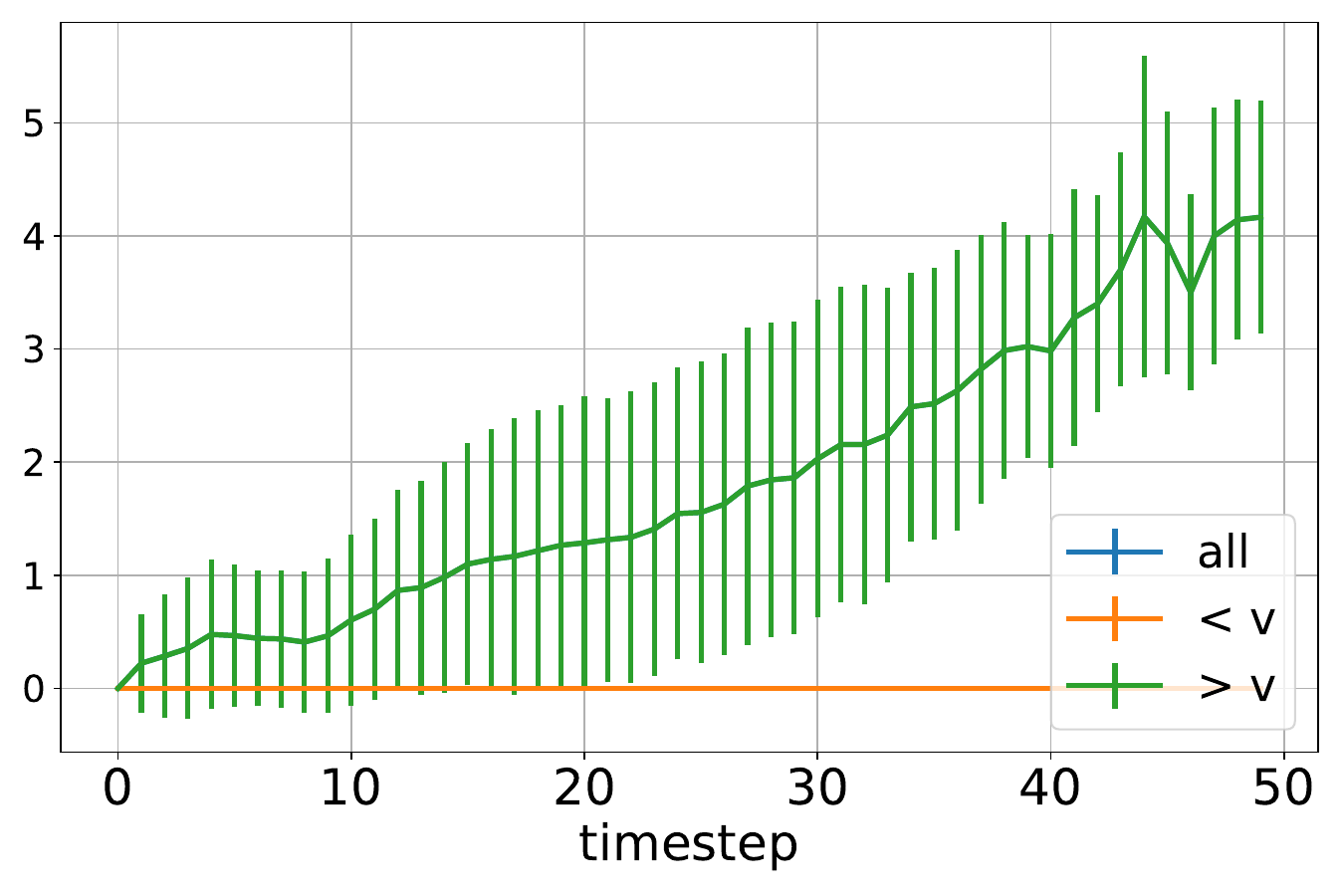}
     \caption{Test 1}
\end{subfigure}
\begin{subfigure}[c]{0.24\columnwidth}
     \includegraphics[width=\textwidth]{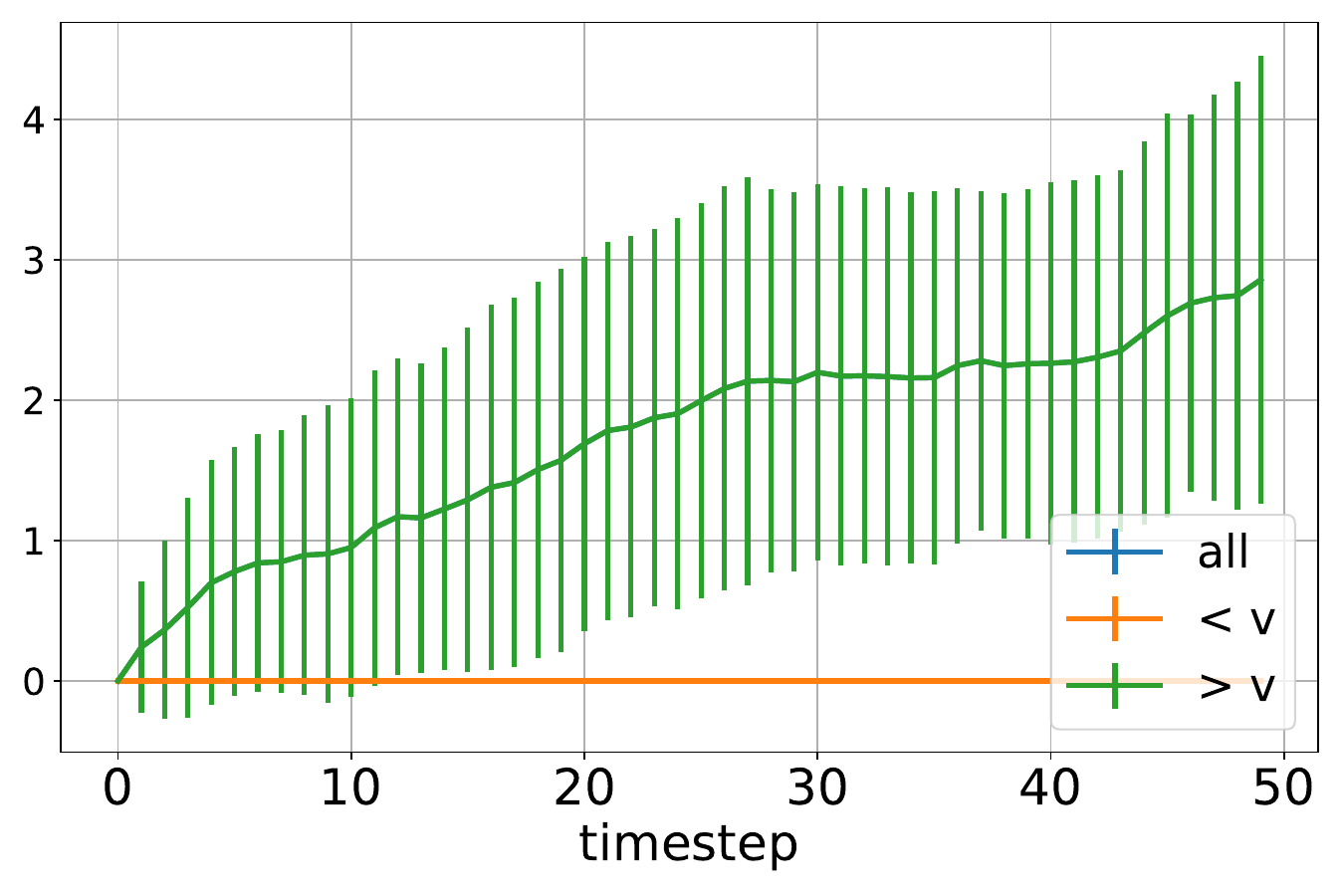}
     \caption{Test 2}
\end{subfigure}
\begin{subfigure}[c]{0.24\columnwidth}
     \includegraphics[width=\textwidth]{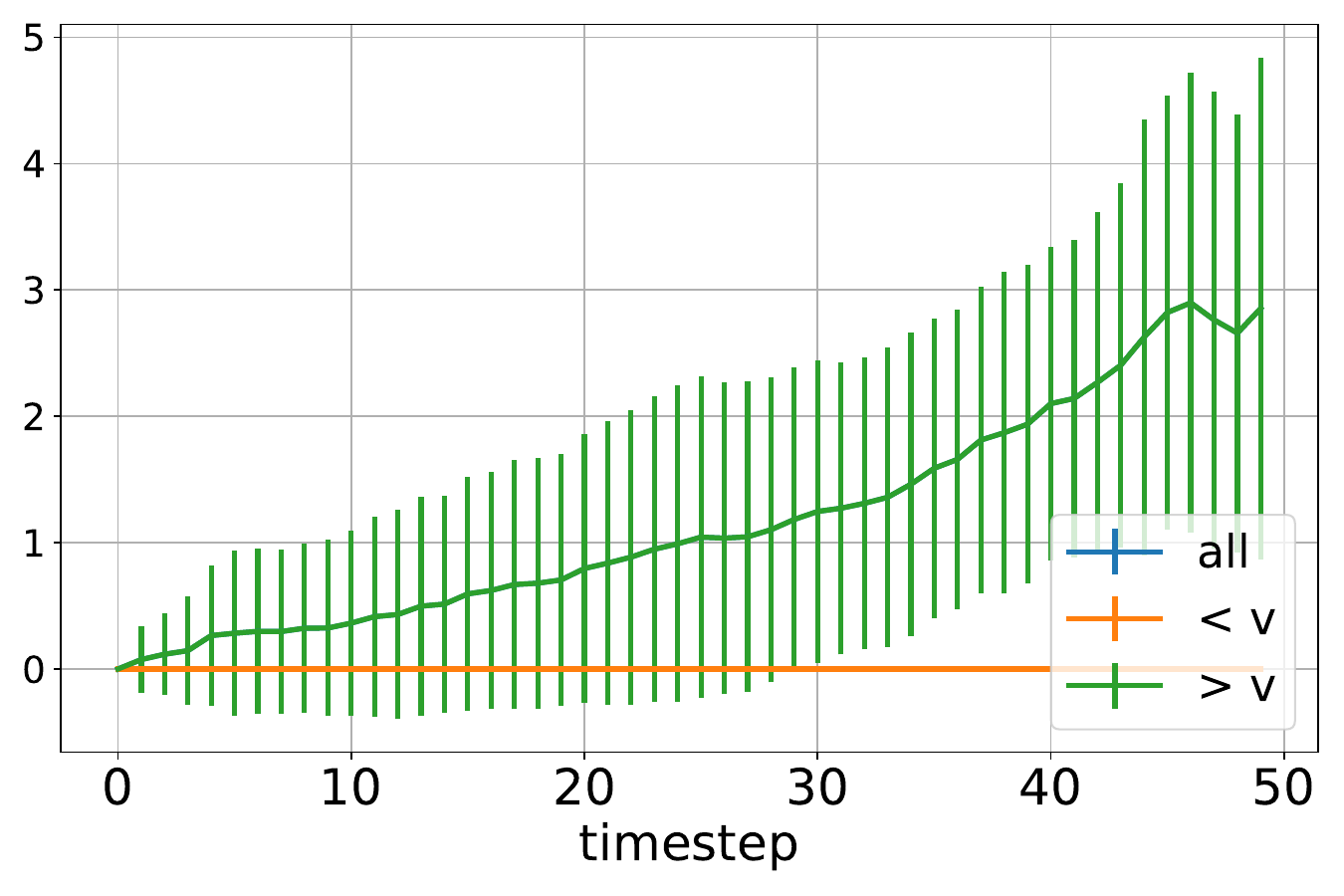}
     \caption{Test 3}
\end{subfigure}
\caption{False positives across timesteps. $<v$ means within the field of view, and $>v$ means outside the field of view.}
\label{fig:fp}
\end{figure}

\end{document}